\documentclass[aps,preprint,showkeys,a4paper,tightenlines,superscriptaddress,12pt]{revtex4}

\usepackage{amsmath}
\usepackage{graphics}
\usepackage{graphicx}
\usepackage{caption}
\usepackage{subcaption}
\usepackage{float}
\usepackage{natbib}
\usepackage[a4paper, total={6in, 8in}]{geometry}

\begin{document}

\title{On the effective surface energy in viscoelastic Hertzian
contacts}
\author{L. Afferrante}
\email{luciano.afferrante@poliba.it}
\address{Department of Mechanics, Mathematics and Management, Polytechnic University
of Bari, Via E. Orabona, 4, 70125, Bari, Italy}

\author{G. Violano}
\address{Department of Mechanics, Mathematics and Management, Polytechnic University
of Bari, Via E. Orabona, 4, 70125, Bari, Italy}

\begin{abstract}
Viscoelasticity and rate-dependent adhesion of soft matter lead to
difficulties in modeling the 'relatively simple' problem of a rigid sphere
in contact with a viscoelastic half-space. For this reason, approximations
in describing surface interactions and viscous dissipation processes are
usually adopted in the literature.

Here, we develop a fully deterministic model in which adhesive interactions
are described by Lennard-Jones potential and the material behaviour with the
standard linear solid model.

Normal loading-unloading cycles are carried out under different driving
conditions. When loading is performed in quasi-static conditions and, hence,
unloading starts from a completely relaxed state of the material, the
effective surface energy is found to monotonically increase with the contact
line velocity up to an asymptotic value reached at high unloading rates.
Such result agrees with existing theories on viscoelastic crack propagation.

If loading and unloading are performed at the same non-zero driving velocity
and, hence, unloading starts from an unrelaxed state of the material, the
trend of the effective surface energy $\Delta \gamma _{\mathrm{eff}}$ with
the contact line velocity is described by a bell-shaped function in a
double-logarithmic plot. The peak of $\Delta \gamma _{\mathrm{eff}}$\ is
found at a contact line velocity smaller than that makes maximum the tangent
loss of the viscoelastic modulus. Furthermore, we show Gent\&Schultz
assumption partly works in this case as viscous dissipation is no longer
localized along the contact perimeter but it also occurs in the bulk material.
\end{abstract}
\maketitle
\section{Introduction}

The adhesive contact between a rigid sphere and a viscoelastic substrate is
a problem of scientific and technological interest, as adhesion of soft
materials plays a key role in biomedicine \cite{Jeong2015}, soft robotics 
\cite{Shintake2018}, sensors \cite{Yao2020}, tapes \cite{Villey2015} and
printing industries \cite{Meitl2006}. Moreover, the contact mechanics of
soft matter is not yet fully understood, due to its intrinsic
viscoelasticity and rate-dependent adhesive features \cite{ViolanoPart1,ViolanoPart2}.

In the 1960s, Lee \& Radok \cite{Lee1960} investigated the adhesiveless
contact between a rigid sphere and a viscoelastic half-space introducing a
simple method to replace the elastic modulus with the viscoelastic creep
function. However, their solution is exact only for the loading phase. The
solution for the unloading phase was formulated by Ting \cite{Ting1966},
whose method is based on the use of the correspondence principle and Laplace
transforms. Greenwood \cite{Greenwood2010} defined Ting's approach for
finding the contact area and applied load an \textit{algebraically messy }%
procedure. To avoid the use of Ting's repeated integrations, he proposed to
exploit the superposition of an assembly of viscoelastic Boussinesq circular
punch indentations.

In the 1970s, two fundamental theories on adhesion of elastic spheres were
formulated: Johnson, Kendall \& Roberts (JKR) theory \cite{JKR1971} and
Derjaguin, Muller \& Toporov (DMT) theory \cite{DMT1975}. The first is based
on the assumption of infinitely short-range adhesive interactions acting
inside the contact area. The latter instead neglects interactions inside the
contact area (where a Hertzian contact shape is assumed) and considers
long-range adhesive interactions acting outside it. Tabor \cite{Tabor}
showed JKR and DMT theories are the extreme limits of a single theory
parametrized by the parameter $\mu =\left[ R\Delta \gamma ^{2}/\left(
E^{\ast 2}\varepsilon ^{3}\right) \right] ^{1/3}$ (known as Tabor's
parameter), where $R$ is the radius of the sphere, $\Delta \gamma $ is the
surface energy of adhesion, $E^{\ast }$ is the composite elastic modulus and 
$\varepsilon $ is the range of action of van der Waals forces. In fact, DMT
theory works in the limit of $\mu \ll 1$ (i.e., for small and stiff
spheres), while JKR theory applies to $\mu \gg 1$ (i.e., for large and
compliant spheres).

More recently, Maugis \cite{Maugis1992}, by exploiting the Dugdale
approximation \cite{Dugdale}, proposed an analytical model based on the
assumption to represent the surface force in terms of a Dugdale cohesive
zone approximation. In this model (known as Maugis-Dugdale (MD) model) the
work of adhesion is assumed to be $\Delta \gamma =\sigma _{0}d_{0}$, where $%
\sigma _{0}$ is the maximum force predicted by the Lennard-Jones force law
and $d_{0}$ is the separation at which the areas under the Dugdale and
Lennard-Jones curves match. JKR, DMT and MD\ theories are derived in
quasi-static conditions and, therefore, cannot be applied to viscoelastic
solids exhibiting rate-dependent adhesion \cite{ViolanoPart1}. Real soft
materials, like natural rubbers and synthetic polymers, seldom show a linear
elastic constituive behaviour. In fact, the contact mechanical response of
soft matter depends on the instantaneous value of its viscoelastic modulus,
which is a function of the temperature and the frequency of excitation.

Starting from the 1980s, different strategies for studying adhesion of
viscoelastic solids have been derived in the framework of classical fracture
mechanics \cite{MB1980,GreenJohn1981,Muller1999}. Gent \& Schultz (GS) \cite%
{GS1972} and Maugis \& Barquins (MB) \cite{MB1980} showed, with experiments, the apparent surface energy depends on the contact line velocity. Such dependence was also predicted by Schapery in his theoretical studies on crack's
propagation in viscoelastic media \cite{Schapery1975a}. The solutions
developed in Refs. \cite{MB1980,GreenJohn1981,Muller1999} moved from GS\
assumption that predicts viscous dissipation is localized along the contact
perimeter, which can be seen as the tip of a circular crack. However, such
assumption requires that the material behaves elastically, which is not
always the case.

In 2000s, Haiat et. al \cite{Haiat2003} formulated a generalization of
Ting's adhesiveless solution by extending the restricted self-consistent
adhesive model by Tabor \cite{Tabor} to viscoelastic spheres. This model
requires to solve simultaneously two equations describing the
time-dependence of the contact and adhesive zones, respectively. For complex
contact histories, nested time integration may lead to decreasing
feasibility of the method. Moreover, adhesive interactions are modeled
according to a double-Hertz approximation \cite{G-J1998}. An alternative approach has been proposed by Lin \& Hui \cite{Lin2002}, who performed
Finite Element (FE) simulations of the adhesive contact between viscoelastic
spheres. In their model, adhesion is implemented by means of a Dugdale cohesive zone model.
Furthermore, Barthel \& Fretigny \cite{BF2009} have proposed theoretical expressions to estimate the impact of viscoelasticity on effective surface energy.
More recently, Sukhomlinov \& M\"{u}ser have extended the original Green's Function Molecular Dynamics approach \cite{Campana&Muser} to investigate the viscous dissipation caused by rough indenters in sliding motion \cite{Sukhomlinov&Muser}.

The depicted state of the art highlights the absence of a model that is able (i) to describe the real physics of the adhesive interactions at the
contact interface, and (ii) to capture rate-dependent adhesion of soft matter, without any assumptions on the location of viscous dissipation. To
achieve these manifold objectives, we herein develop a fully deterministic
Finite Element model aimed at investigating the viscoelasticity induced
adhesion enhancement in the contact of a spherical indenter with a smooth
flat substrate. In our model, adhesion is described by means of non-linear
springs following the Lennard-Jones traction-gap law. We study the
enhancement of the effective surface energy with the contact line velocity
during retraction. Numerical results are compared with the predictions of
the crack propagation models of Greenwood \cite{Greenwood2004}, and Persson
\& Brenner (PB) \cite{PB2005}, which both show to be accurate. We also
discuss the validity of GS\ assumption and show when viscous dissipation may
occur in the bulk material.

\section{Formulation}

\begin{figure}[tbp]
\begin{center}
\includegraphics[trim=0.8cm 9.5cm 0.8cm 9.5cm,clip=true,width=12.0cm]{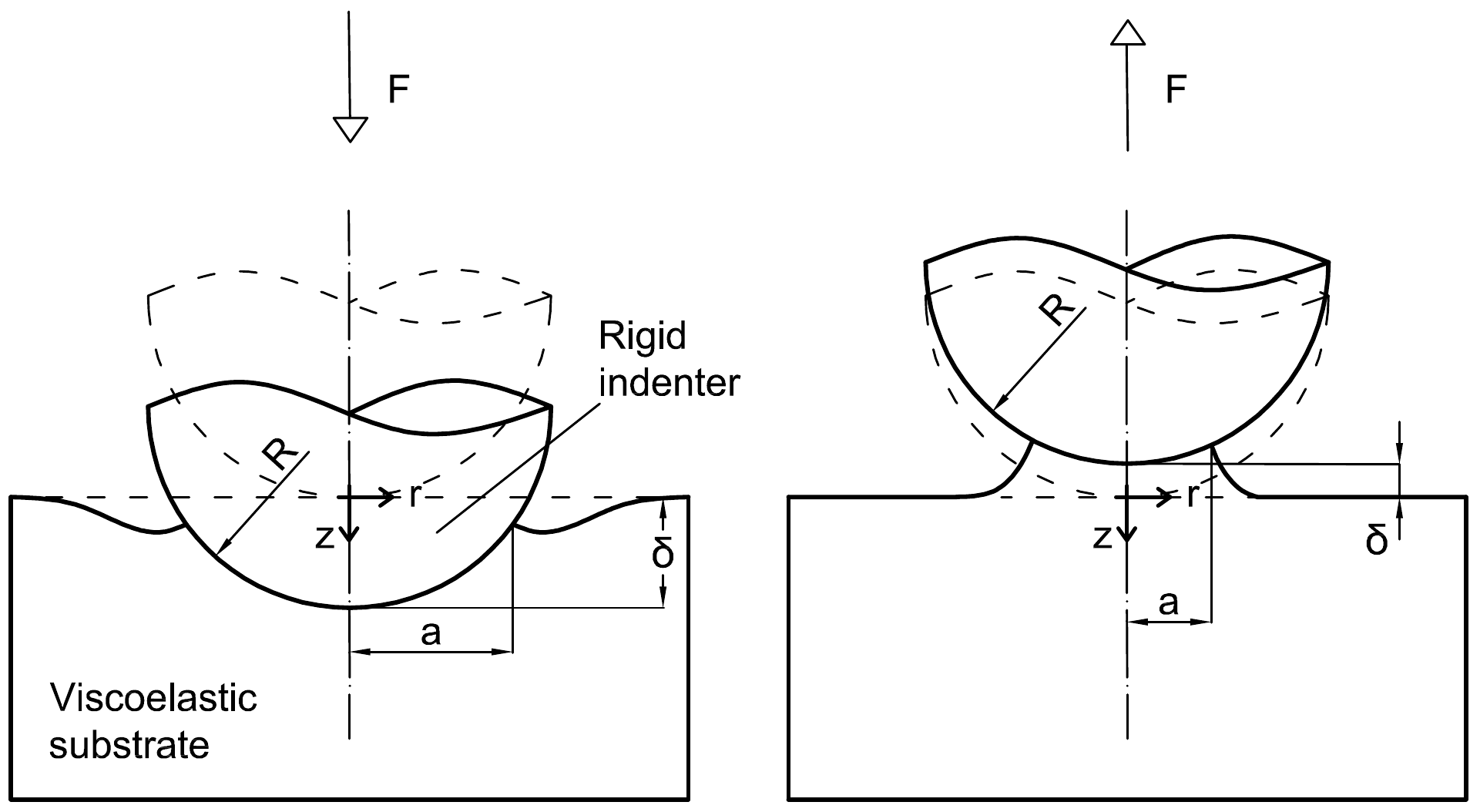}
\end{center}
\caption{The problem under investigation: a rigid hemisphere of radius $R$
is pressed into a viscoelastic half-space and then pulled apart from it.}
\label{fig1}
\end{figure}

Figure \textbf{\ref{fig1}} gives a graphical representation of the problem under
investigation: a rigid hemisphere of radius $R$ is pressed into a
viscoelastic half-space and then pulled apart from it. The normal driving
velocity $V=d\delta /dt$ is assumed constant during the loading and
unloading, being $\delta $ the contact penetration. Notice $\delta $ is
measured at $r=0$ and corresponds to the normal displacement of the surface
of the half-space with respect to its undeformed configuration (black dashed
line). We denoted with $a$ and $F$, the contact radius and applied load,
respectively, which is assumed positive when compressive.

\begin{figure}
\centering
     \begin{subfigure}[b]{0.8\textwidth}
        \includegraphics[width=12.0cm]{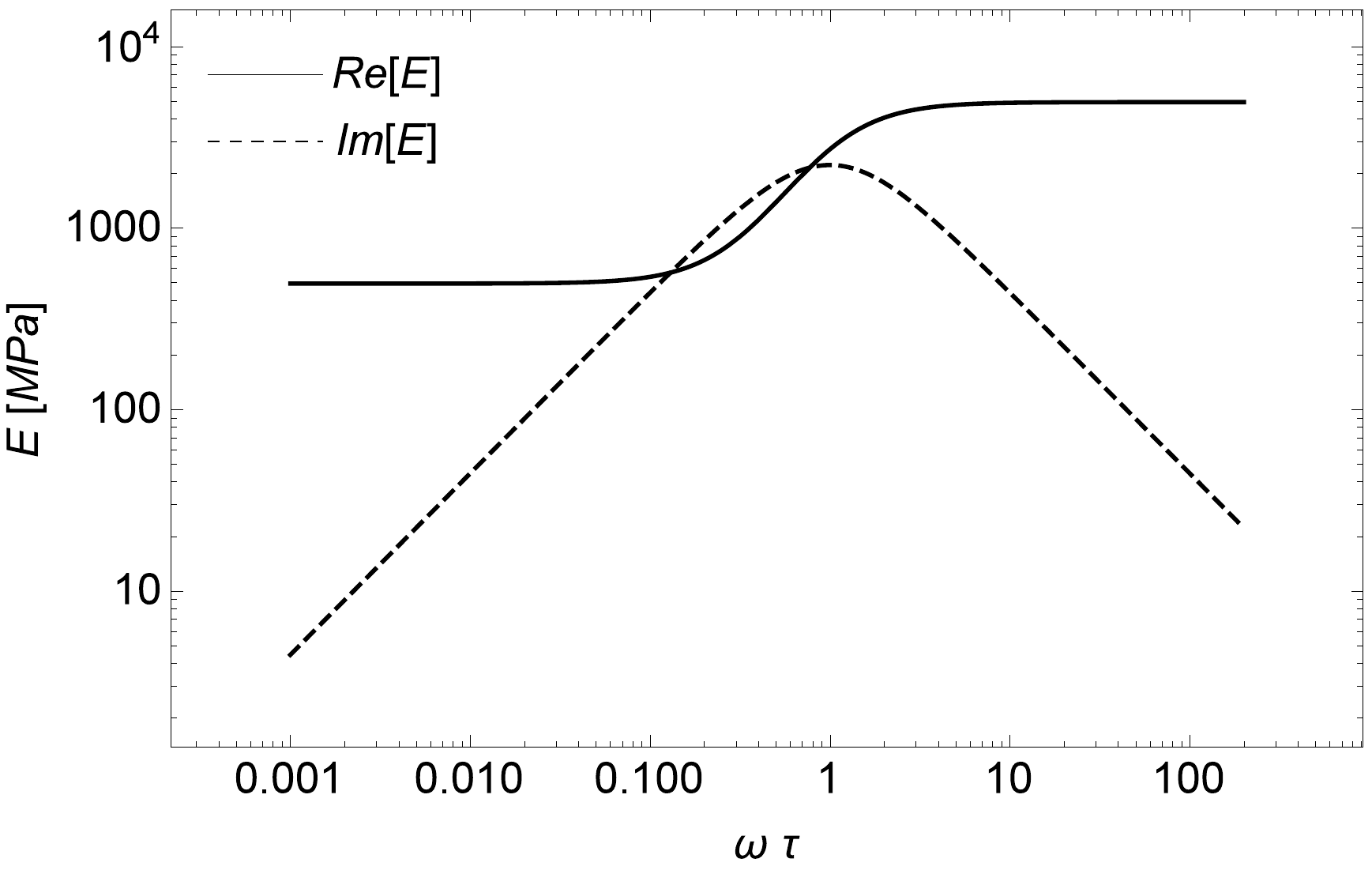}
         \caption{}
     \end{subfigure}\\
     \begin{subfigure}[b]{0.8\textwidth}
         \includegraphics[width=11.5cm]{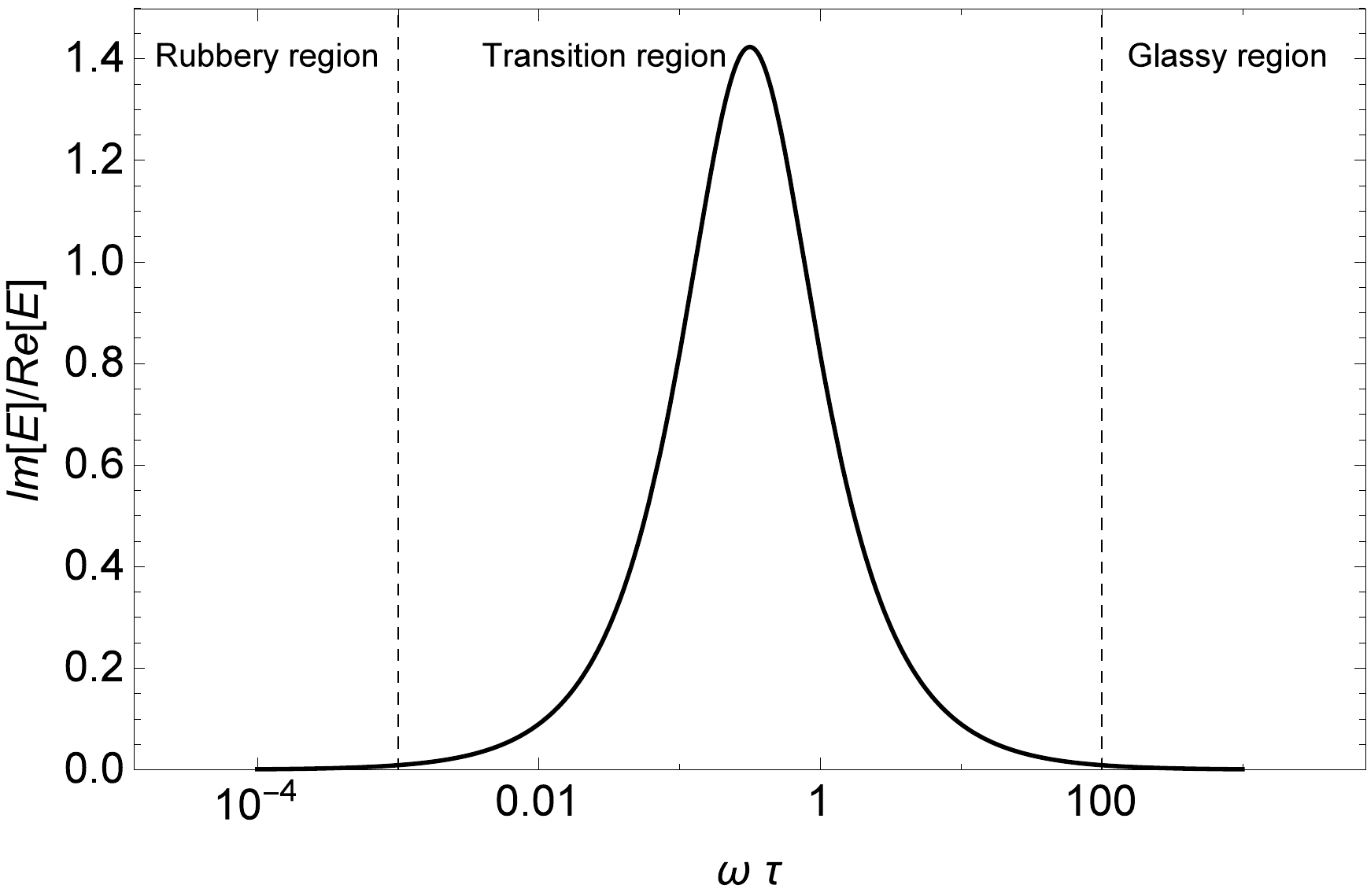}
         \caption{}
     \end{subfigure}\\
     \caption{(a) The real (solid line) and imaginary (dashed line) parts of the viscoelastic modulus $E(\omega )$ for a standard linear viscoelastic solid. $Re[E]$ and $Im[E]$ are shown in terms of the dimensionless frequency $\omega \tau $. The curves are given for $E(0)/E(\infty )=0.1$. (b) The dependence of the loss tangent $Im[E]/Re[E]$ on the dimensionless frequency $\omega \tau $.}
     \label{fig2}
\end{figure}

A standard linear solid is used to describe viscoelasticity in the
half-space. Figure \textbf{\ref{fig2}} shows the dependence of the
viscoelastic modulus $E$ on the frequency $\omega $. Specifically, Fig. 
\textbf{\ref{fig2}a} shows the variation with the dimensionless frequency $%
\omega \tau $ of the real and imaginary parts of $E$, while Fig. \textbf{%
\ref{fig2}b} the variation of the loss tangent $Im[E]/Re[E]$ with $\omega 
\tau $.

At low frequencies, the material is in the `rubbery' region where the real
part $Re[E]$ of the viscoelastic modulus is constant and the viscous
dissipation related to the imaginary part $Im[E]$ is negligible (loss
tangent $Im[E]/Re[E]\approx 0$). At high frequencies, i.e., in the `glassy'
region, viscous dissipation is again negligible and the real part of $E$ is
constant but much larger than in the `rubbery' region. Viscous dissipation
occurs at intermediate frequencies in the `transition' region where the loss
tangent takes non-zero values.

\subsection{Numerical model}

\begin{figure}[tbp]
\begin{center}
\includegraphics[trim=0.5cm 10.5cm 0.5cm 10.5cm,clip=true,width=12.0cm]{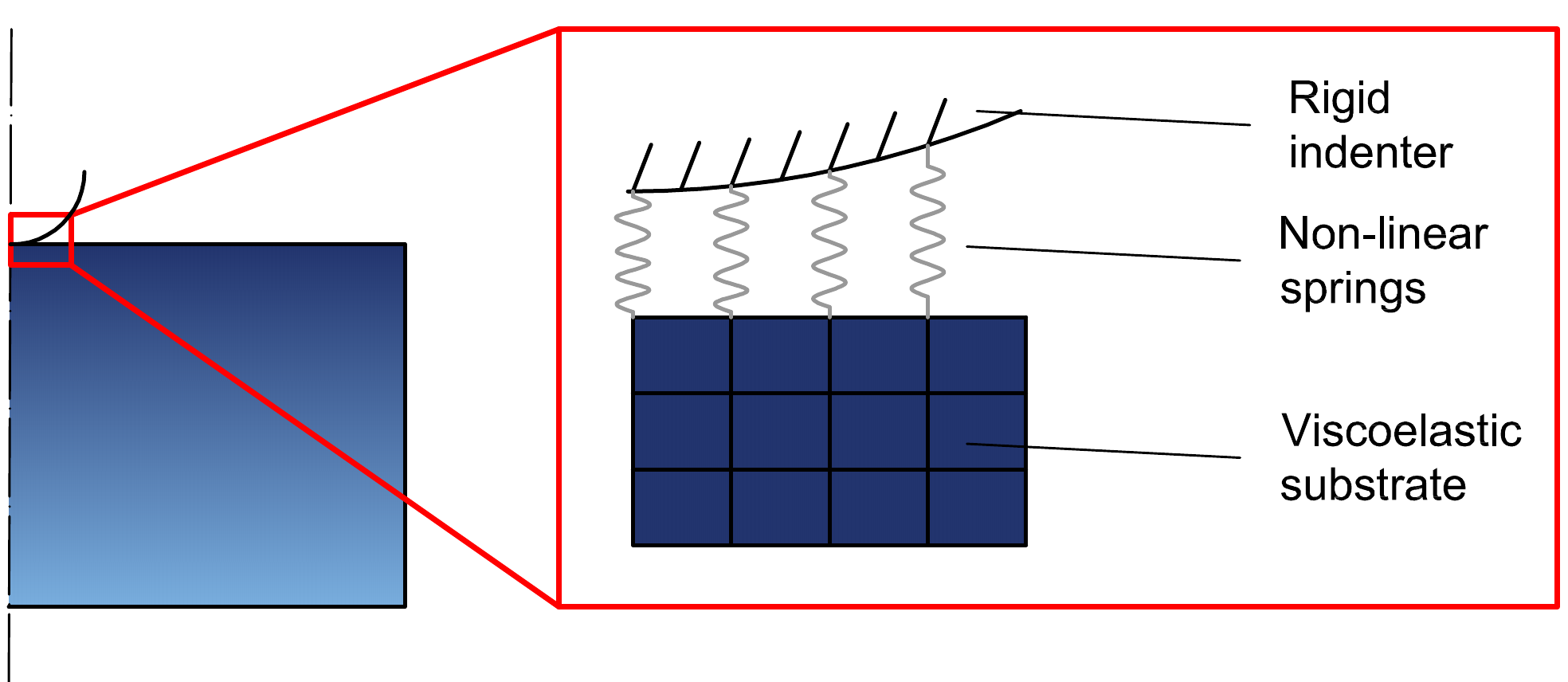}
\end{center}
\caption{Graphical representation of the finite element model.}
\label{fig3}
\end{figure}

A Finite Element (FE) model (Fig. \textbf{\ref{fig3}}) has been developed to study the problem
sketched in Fig. \textbf{\ref{fig1}}. The substrate has been modeled with axisymmetric plane
elements with linear shape functions. Constraints are applied at the bottom
boundary. For the rigid sphere we need to define only nodes on the surface
where a single master node has been identified. On this node force and
displacement are applied and its degree of freedoms are coupled with those
of all other sphere nodes by simple constraint equations.

The adhesive interaction has been modeled by nonlinear spring elements with
a traction-displacement relation according to the traction-gap law of
Lennard-Jones (L-J)%
\begin{equation}
p\left( r\right) =\frac{8\Delta \gamma }{3\varepsilon }\left[ \left( \frac{%
\varepsilon }{g\left( r\right) }\right) ^{3}-\left( \frac{\varepsilon }{%
g\left( r\right) }\right) ^{9}\right] \text{.}
\end{equation}

Notice the compression (elongation) of the springs is the sum of the total
'rigid' displacement $\Delta $ of the sphere and the surface displacement $u$
of the deformable substrate%
\begin{equation}
g_{0}\left( r\right) -g\left( r\right) =\Delta +u\left( r\right)
\end{equation}%
being $g_{0}=h_{0}+r^{2}/\left( 2R\right) $ the initial gap function, where $%
h_{0}$ is the assumed initial distance between the sphere and the substrate
at $r=0$.

This type of approach was already adopted in previous works (see, for
example, Refs. \cite{Kadinetal2008, SongKomvo2011}) to investigate similar
problems with linear elastic and/or elasto-plastic materials.

As a result, the adhesive force exerted by a spring connected to a surface
node at distance $r$ from the contact center is $dF\left( r\right) =p\left(
r\right) 2\pi rdr$, except for the central node where we have $dF\left(
r\right) =p\left( r\right) \pi dr^{2}$.

The substrate has been modeled as a linear viscoelastic material.
Specifically, a classical linear standard model was used with a single
relaxation time $\tau $ and a Prony series representation was adopted%
\begin{equation}
E\left( t\right) =E_{0}+\left( E_{\infty }-E_{0}\right) \exp \left( -t/\tau
\right)
\end{equation}%
where $E_{0}$ is the elastic modulus at zero frequency and $E_{\infty }$ is
the elastic modulus at high frequency.

The stress $\mathbf{\sigma }$ is calculated according to the following
constitutive equation%
\begin{equation}
\mathbf{\sigma =}\int_{0}^{t}E\left( t-t^{\prime }\right) \frac{d\mathbf{%
\varepsilon }}{dt^{\prime }}dt^{\prime }  \label{s}
\end{equation}%
where $\mathbf{\varepsilon }$ is the strain and $E$ the relaxation function.

To calculate stresses at time $t_{i+1}=t_{i}+\Delta t$, the classical
incremental finite element procedure has been used%
\begin{equation}
\mathbf{\sigma }\left( t_{i+1}\right) =\mathbf{\sigma }\left( t_{i}\right)
\exp \left( -\Delta t/\tau \right) +\int_{t_{i}}^{t_{i+1}}\left( E_{\infty
}-E_{0}\right) \exp \left( -\frac{t_{i+1}-t^{\prime }}{\tau }\right) \frac{d%
\mathbf{\varepsilon }}{dt^{\prime }}dt^{\prime }.
\end{equation}

Finally, the dissipated energy in the viscoelastic substrate is calculated
as the energy required to deform the dashpot in the Maxwell element.
Therefore, the increment of energy dissipated over a time increment $\Delta t
$ is%
\begin{equation}
\Delta E_{d}=\mathbf{\sigma }\cdot \Delta \mathbf{\varepsilon }%
_{dashpot}\Delta t
\end{equation}%
where $\Delta \mathbf{\varepsilon }_{dashpot}$ is the dashpot strain
increment over time $\Delta t$ given by%
\begin{equation}
\Delta \mathbf{\varepsilon }_{dashpot}=\Delta \mathbf{\varepsilon }-\alpha
\Delta \mathbf{\sigma }
\end{equation}%
being $\alpha $ the Maxwell spring compliance.

The prepared code was solved with the aid of ANSYS solver.

\section{Results and Discussion}

All results presented in this section are obtained for $E_{0}/E_{\infty
}=0.1 $ and $\mu _{0}=\left[ R\Delta \gamma ^{2}/\left( E_{0}^{2}\varepsilon
^{3}\right) \right] ^{1/3}\approx 3.85$ and are given in terms of the
following dimensionless quantities: $\hat{V}=V\tau /\varepsilon $, $\hat{a}%
=a/R$, $\hat{\delta}=\delta /\varepsilon $, $\hat{F}=F/\left( 1.5\pi R\Delta
\gamma \right) $, and $\hat{p}=p\varepsilon /\Delta \gamma $.

\begin{figure}
\centering
     \begin{subfigure}[b]{0.8\textwidth}
        \includegraphics[width=12.0cm]{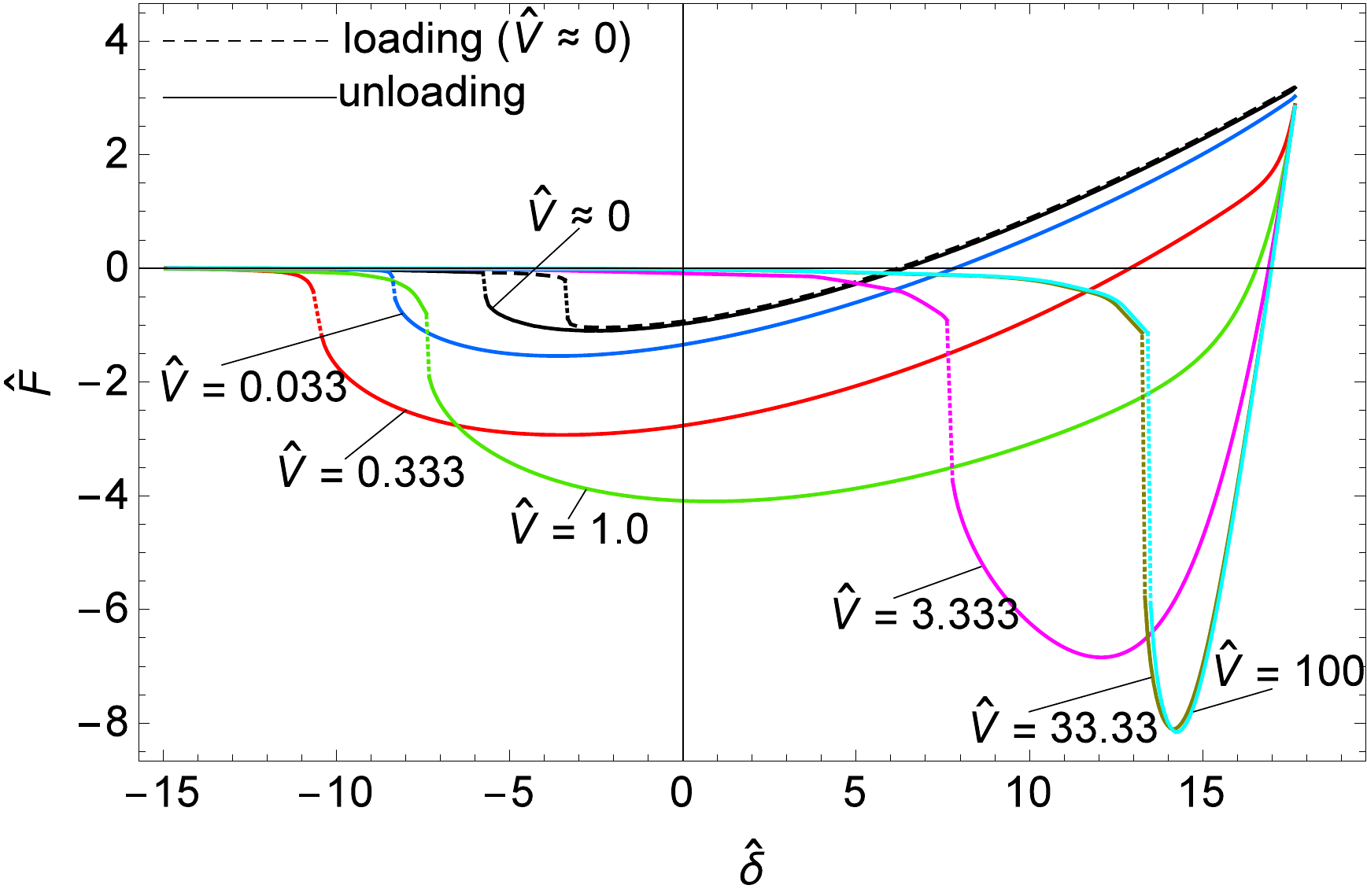}
         \caption{}
     \end{subfigure}\\
     \begin{subfigure}[b]{0.8\textwidth}
         \includegraphics[width=12.0cm]{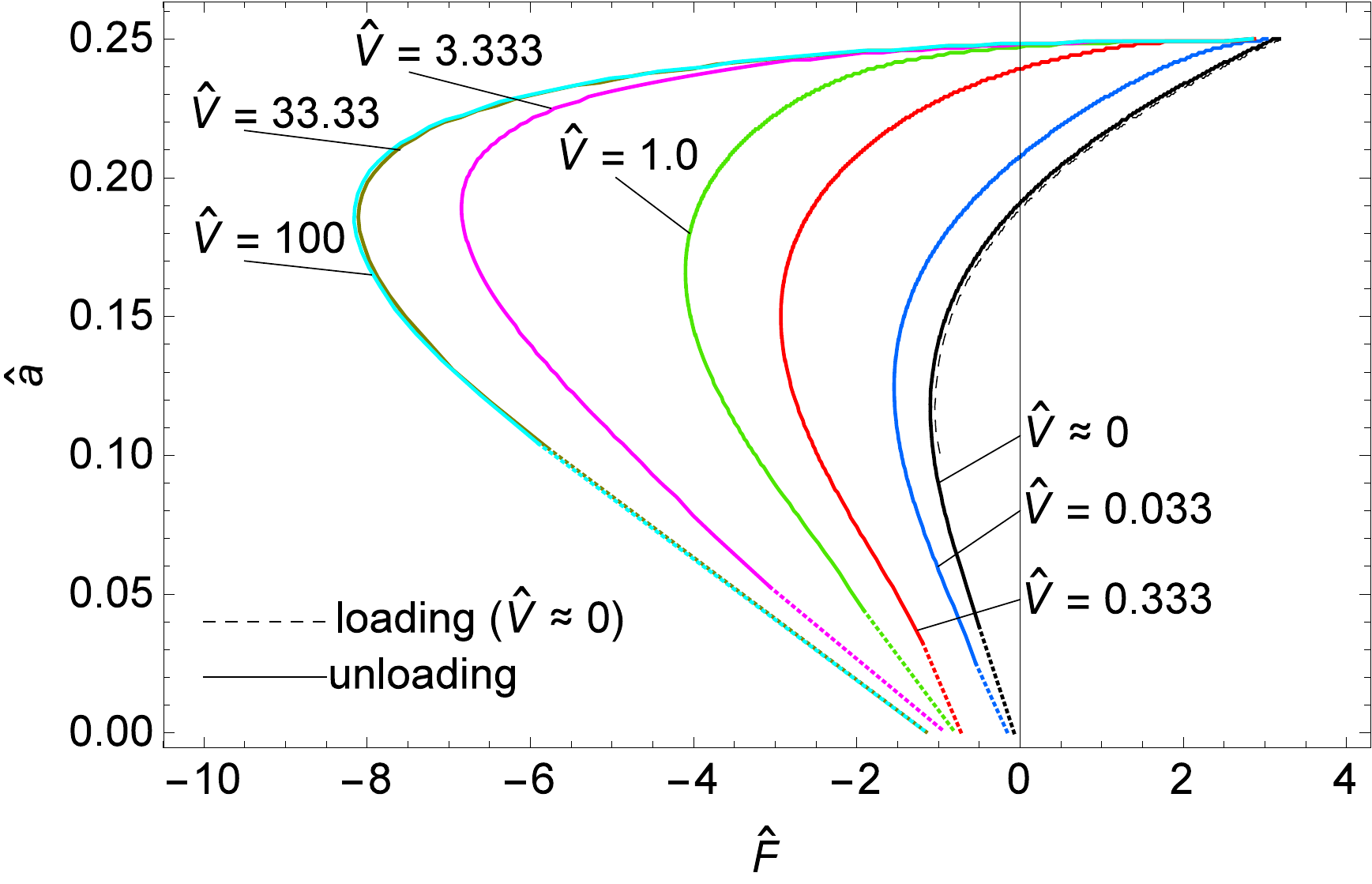}
         \caption{}
     \end{subfigure}\\
     \caption{(a) The dimensionless force $\hat{F}$ as a function of the dimensionless imposed
approach $\hat{\delta}$. Loading (dashed line) is performed at vanishing
velocity ($V\approx 0$) so one can neglect viscous effects. Unloading
(coloured solid lines) is instead performed at different driving velocities $%
\hat{V}$. Dotted lines connect the two branches of the curves where jump-in
(loading phase) and jump-off (unloading phase) instabilities occur. Results
are given for a standard linear solid with $E_{0}/E_{\infty
}=0.1$ and $\protect\mu _{0}=[R\Delta \protect\gamma ^{2}/(E_{0}^{2}%
\protect\varepsilon ^{3})]^{1/3}\approx 3.85$.  (b) The dimensionless contact
radius $\hat{a}$ as a function of the dimensionless force $\hat{F}$. Loading
(dashed line) is performed at vanishing velocity ($V\approx 0$) so one can
neglect viscous effects. Unloading (coloured solid lines) is instead
performed at different driving velocities $\hat{V}$. Dotted lines identify
the regions where jump-off instabilities occur. Notice no perceptible
difference between the curves can be identified for $\hat{V}>33.33$. Results
are given for a standard linear solid with $E_{0}/E_{\infty
}=0.1$ and $\protect\mu _{0}=[R\Delta \protect\gamma ^{2}/(E_{0}^{2}%
\protect\varepsilon ^{3})]^{1/3}\approx 3.85$.}
     \label{fig4}
\end{figure}

A first set of numerical simulations has been run at vanishing approaching
velocity ($V\approx 0$) to neglect time dependent effects during the loading
phase. In this way, we are sure that the material is in a strain relaxed state at the end of loading and the following unloading phase always
starts from the same initial conditions corresponding to the `rubbery'
region of Fig. \textbf{\ref{fig2}}. Figure \textbf{\ref{fig4}} shows the resulting
curves after a cycle of loading-unloading performed at different pulling
velocities. Specifically, with reference to Fig. \textbf{\ref{fig4}a}, hysteresis
dissipation (given by the difference between the work spent to bring the
bodies in contact and that required to separate them) occurs even when the
pulling velocity is negligible (i.e., in absence of viscous effects). In
such case, energy loss (quantitatively given by the area between the loading
and unloading curves) is exclusively due to the adhesive hysteresis related to the jumping-on and jumping-off phenomena (dotted lines identify the points where jump-on and jump-off instabilities occur). In this respect, Refs. \cite{Wangetal2021, ViolAff2021} have proposed a comprehensive investigation of the elastic adhesive hysteresis occurring between approach and retraction.

As expected, increasing the pulling velocity, dissipation increases and the
unloading curves (coloured solid lines) more and more deviate from the
loading one (dashed line). As a result, the pull-off force $F_{\mathrm{PO}}$
(i.e., the maximum tensile force) is observed to monotonically increase. It
reaches a maximum value at about $\hat{V}\approx 33.33$, after which it
remains constant. Moreover, notice that at the highest velocities
snap-off occurs at positive (compressive) penetrations $\delta $ because
increasing the pulling velocity the time for the viscoelastic substrate to
`relax' and recovery its deformation reduces. As a result, the sphere
imprint on the substrate can be clearly observed even when the sphere has
been completely detached.

Figure \textbf{\ref{fig4}b} shows the dependence of the contact radius $a$ on the
applied force during the loading and unloading simulations. In the present
context, a rigorous definition of the contact radius can be disputable
because Lennard-Jones force law imposes the gap between contacting
surfaces is always nonzero. As observed by Feng in Ref. \cite{Feng2001},
where he investigated the adhesive contact between a rigid half-space and a
deformable elastic sphere, the edge of the flattened area coincides well
with the point where the tensile stress reaches its peak when the Tabor
parameter $%
%TCIMACRO{\U{3bc} }%
%BeginExpansion
\mu
%EndExpansion
>1$. This agrees with previous observations of Greenwood \cite{Greenwood1997}
who proposed to identify the edge of contact area at the point where the
tensile stress takes its maximum value. However, for smaller $%
%TCIMACRO{\U{3bc} }%
%BeginExpansion
\mu
%EndExpansion
$, the edge of the contact area is less easily identified and seems to
coincide with the point where the local Lennard-Jones pressure becomes zero 
\cite{Feng2001}. In our calculations, we find a perfect coincidence of the
deformed profile of the substrate with that of the Hertzian rigid indenter
up to the point where the tensile stress is maximum. Therefore, in the
present calculations, the contact radius is identified as the radius at
which the pressure takes its maximum tensile value. No perceptible
difference between the curves can be identified for $\hat{V}>33.33$. Also,
during the initial phase of unloading, the contact radius is practically
constant. Such period (identified as stick time in Refs. \cite%
{Haiat2003,ViolanoPart1,ViolFront}) increases with the speed and reaches a
maximum at about $\hat{V}=33.33$, above which no further variation is
observed.

The pull-off process of a sphere from a substrate is analogous to the
opening of a circular crack. In cracks propagation, viscoelastic dissipation
is known to determine an increase in the propagation energy $\Delta \gamma
_{\mathrm{eff}}$ \cite{Schapery1975a,Greenwood2004,PB2005}. It depends on
the nature of the processes occurring in the crack-tip process zone, which
is the region close to the crack tip where the classical Irwin equations 
\cite{Irwin1968} relating stresses to the distance from the crack tip are no
longer valid.

It has been shown experimentally \cite{MB1980,GS1972} that $\Delta \gamma
_{\mathrm{eff}}$ depends on the crack tip velocity $V_{\mathrm{c}}$ and the
temperature $T$ according to%
\begin{equation}
\Delta \gamma _{\mathrm{eff}}=\Delta \gamma \lbrack 1+f(a_{\mathrm{T}}V_{%
\mathrm{c}})]  \label{GS}
\end{equation}%
where $a_{\mathrm{T}}$ is the Williams-Landel-Ferry (WLF) \cite{WLF1955}
factor depending on the temperature and $f(a_{\mathrm{T}}V_{\mathrm{c}})$ is
the function of the viscoelastic energy dissipation in front of the crack
tip. This function is characteristic of the viscoelastic material and is
independent of the contact geometry (see \cite{MB1980}).

In the literature, two different approaches have been developed for studying
crack propagation in viscoelastic solids. The first approach is based on the
cohesive-zone model, which assumes a constant stress $\sigma _{0}$ in the
process zone. Greenwood \cite{Greenwood2004} exploited a Maugis-Dugdale
surface force law to find the dependence of the surface (propagation) energy
on the crack tip velocity. For the standard linear solid used in our work,
one obtains%
\begin{equation}
\Delta \gamma _{\mathrm{eff}}=\Delta \gamma \left[ \frac{E_{0}}{%
E_{\infty }}+\frac{1}{2}\left( 1-\frac{E_{0}}{E_{\infty }}\right) \frac{d}{%
V_{\mathrm{c}}\tau }\frac{E_{0}}{E_{\infty }}\int_{0}^{1}d\xi H(\xi )e^{-%
\frac{d}{V_{\mathrm{c}}\tau }\frac{E_{0}}{E_{\infty }}(1-\xi )}\right] ^{-1}
\label{GW}
\end{equation}%
where%
\begin{equation}
H(\xi )=2\xi ^{1/2}-(1-\xi )\ln \left( \frac{1+\xi ^{1/2}}{1-\xi ^{1/2}}%
\right)
\end{equation}%
and $d=\pi E_{0}\Delta \gamma /(4\sigma _{0}^{2})$ is the width of the
process zone.

The second approach is instead based on energetic equilibrium. It was
developed by Persson \& Brenner (PB) \cite{PB2005}, who showed that the
effective energy required to advance the crack tip by one unit area is
related to the viscoelastic modulus and can be calculated theoretically as%
\begin{equation}
\Delta \gamma _{\mathrm{eff}}=\Delta \gamma \left[ 1-\frac{2E_{0}}{\pi}%
\int_{0}^{2\pi V_{\mathrm{c}}/s}d\omega \frac{F(\omega)}{\omega}Im%
\left( \frac{1}{E(\omega)}\right) \right]^{-1}  
\label{PB}
\end{equation}%
where $F(\omega )=[1-(\omega s/(2\pi V_{\mathrm{c}}))^{2}]^{1/2}$. In
eq. (\ref{PB}), $s=s(V_{\mathrm{c}})$ depends on the crack tip velocity and
can be determined if one assumes that the stress at the crack tip takes some
critical value $\sigma _{\mathrm{c}}$ necessary to break the atomic bonds
for the tip to propagate. This gives%
\begin{equation}
\frac{\Delta \gamma _{\mathrm{eff}}}{\Delta \gamma }=\frac{s}{s_{0}}
\end{equation}%
where $s_{0}$ is the crack tip radius for a very slowly moving crack and
takes values of the order of the nanometer. Specifically, the crack tip
cut-off radius $s_{0}$ in the Persson-Brener model is evaluated as $%
s_{0}=E_{0}\Delta \gamma /\left( 2\pi \sigma _{c}^{2}\right) $, where $%
\sigma _{c}$ is the characteristic stress necessary for bond breaking at the
crack tip. Notice $\sigma _{c}$ is not equivalent to $\sigma _{0}$ in the
cohesive-zone approach as instead erroneously assumed in recent papers \cite%
{ciava, ciavaMcMeeking}.

The mechanism regulating the contact opening occurring during the unloading
phase in our numerical simulations is the same as the crack propagation at
the interface between a viscoelastic solid and a rigid countersurface \cite%
{Persson2021}. Maugis \& Barquins \cite{MB1980} and Charmet et al.\cite%
{Charmet1999} observed that viscoelastic losses are localized at the crack
tip and bulk displacements are purely elastic. Therefore, a first estimate
of the increase in the effective surface energy $\Delta \gamma _{\mathrm{eff}%
}$ due to viscoelastic dissipation in the vicinity of the front of
detachment can be done according to (see \cite{Lorenz2013})%
\begin{equation}
\frac{\Delta \gamma _{\mathrm{eff}}}{\Delta \gamma }\approx \frac{F_{%
\mathrm{PO}}}{F_{\mathrm{PO,}V_{\mathrm{c}}=0}}
\end{equation}%
where $F_{\mathrm{PO,}V_{\mathrm{c}}=0}$ is the pull-off force measured in
the quasi-static limit ($V\approx 0$).

\begin{figure}[tbp]
\begin{center}
\includegraphics[width=12.0cm]{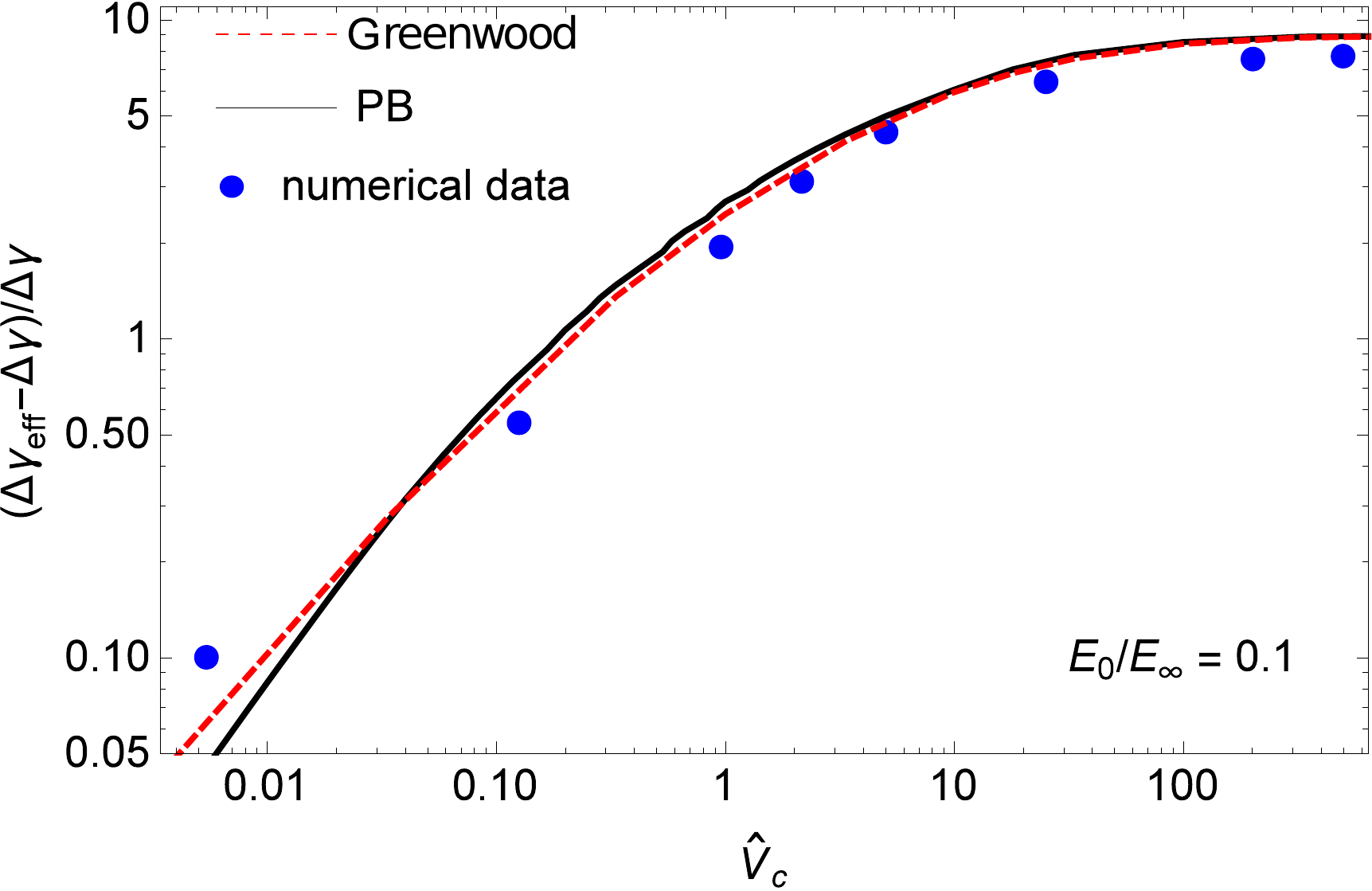}
\end{center}
\caption{Relative increase of the interfacial viscoelastic energy $(\Delta 
\protect\gamma _{\mathrm{eff}}-\Delta \protect\gamma )/\Delta \protect\gamma 
$ as a function of the dimensionless contact line velocity $\hat{V}_{\mathrm{%
c}}$ measured at snap-off. Results are shown in a double logarithmic plot.
The red dashed line is the Greenwood's solution (\protect\ref{GW}) obtained
with $\sigma _{0}\approx 1.026\Delta \gamma /\varepsilon $. Solid line
refers instead to the prediction given by Persson-Brener (PB) equation (%
\protect\ref{PB}) with $s_{0}/\varepsilon \approx 3.33$, i.e., $s_{0}=1$ $%
\mathrm{nm}$ for $\varepsilon =0.3$ $\mathrm{nm}$. Data are given for a standard
linear solid with $E_{0}/E_{\infty }=0.1$ and $\protect\mu %
_{0}=[R\Delta \protect\gamma ^{2}/(E_{0}^{2}\protect\varepsilon %
^{3})]^{1/3}\approx 3.85$.}
\label{fig5}
\end{figure}

Figure \textbf{\ref{fig5}} shows, in a double logarithmic plot, the relative increase
of the effective surface energy $\Delta \gamma _{\mathrm{eff}}$ in terms
of the detachment front propagation velocity $V_{\mathrm{c}}=-da/dt$.
Specifically, $V_{\mathrm{c}}$ is estimated at pull-off (i.e., at the
instant where the tensile force is maximum). Greenwood's solution (red dashed
line) and PB's one (black solid line) are also shown as comparison. The first is obtained with $\sigma _{0}$ equal to the maximum stress of
Lennard-Jones law ($\sigma _{0}\approx 1.026\Delta \gamma /\varepsilon $%
), while the latter assuming $s_{0}/\varepsilon \approx 3.33$ (i.e., $s_{0}=1$%
\textrm{\ nm} for $\varepsilon =0.3$ \textrm{nm}).

As expected (see Refs. \cite{GreenJohn1981,Greenwood2004,PB2005}), $\Delta
\gamma _{\mathrm{eff}}\approx \Delta \gamma E_{\infty }/E_{0}\gg \Delta
\gamma $ for high pulling velocities.

\begin{figure}[tbp]
\begin{center}
\includegraphics[width=12.0cm]{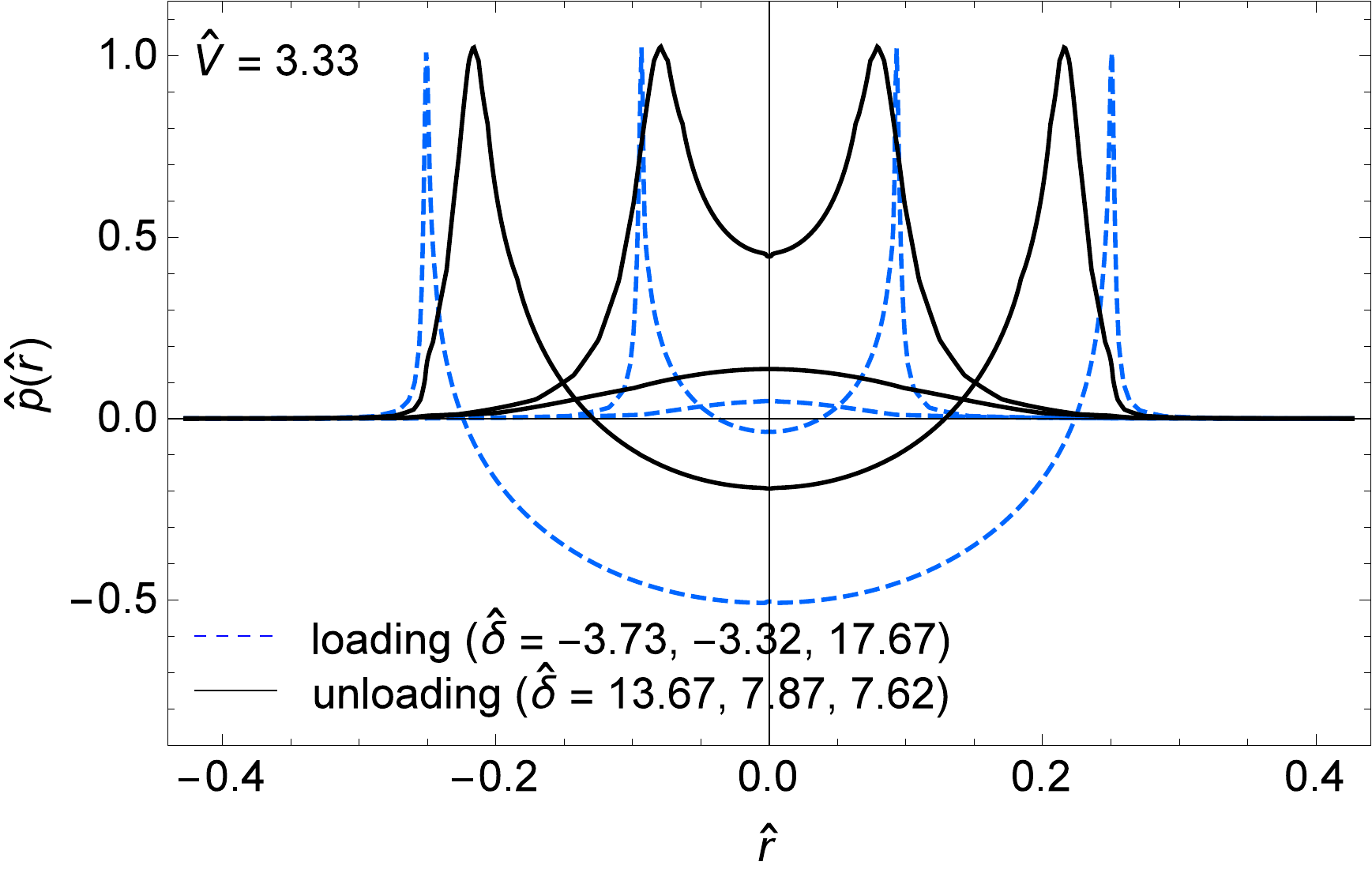}
\end{center}
\caption{The dimensionless contact pressure ($\hat{p}=p\varepsilon /\Delta 
\gamma $) distribution on the contact region for different values of the dimensionless
imposed approach $\hat{\delta}$ and dimensionless driving
velocity $\hat{V}=3.33$. Results are given for a standard linear solid with $E_{0}/E_{\infty }=0.1$ and $\protect\mu %
_{0}=[R\Delta \protect\gamma ^{2}/(E_{0}^{2}\protect\varepsilon %
^{3})]^{1/3}\approx 3.85$.}
\label{fig6}
\end{figure}

Figure \textbf{\ref{fig6}} shows the contact pressure distribution on the contact
region for different values of the approach $\delta $. Dashed lines refer to
the loading phase, while solid lines to the unloading one. The occurrence of
positive tractions is clearly observable at the edges of the contact area
when contact is established.

\begin{figure}
\centering
     \begin{subfigure}[b]{0.8\textwidth}
        \includegraphics[width=12.0cm]{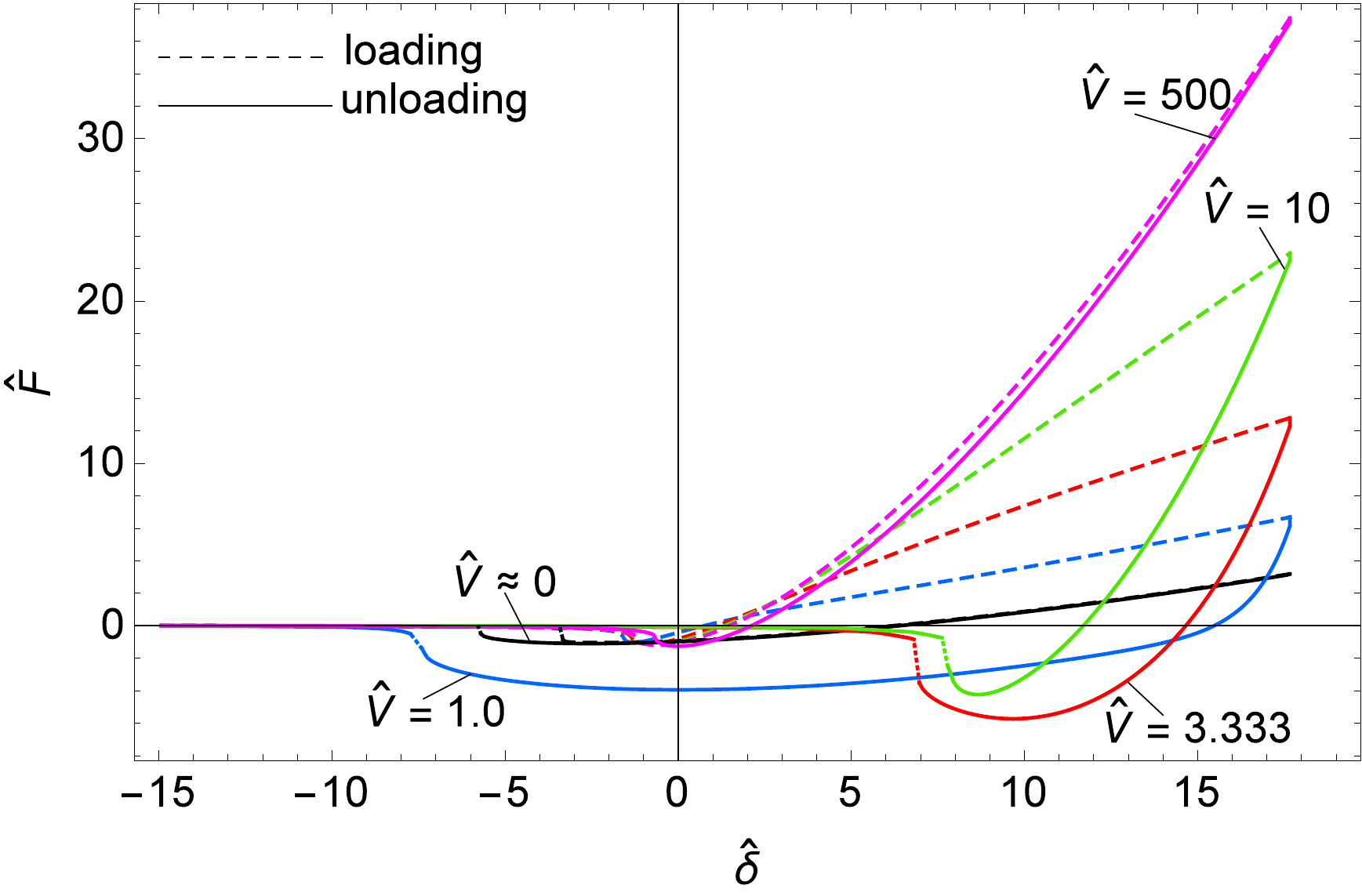}
         \caption{}
     \end{subfigure}\\
     \begin{subfigure}[b]{0.8\textwidth}
         \includegraphics[width=12.0cm]{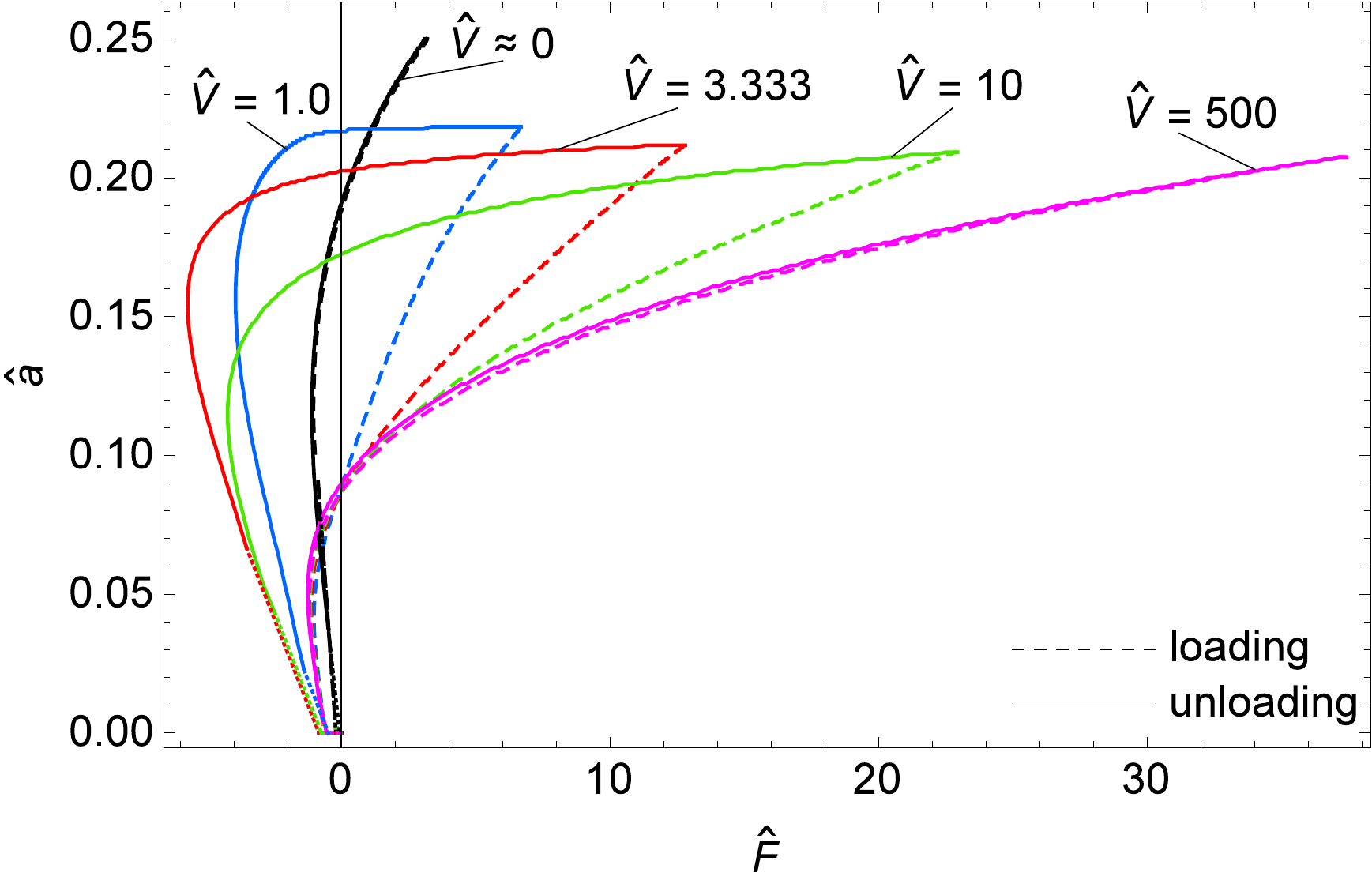}
         \caption{}
     \end{subfigure}\\
     \caption{(a) The dimensionless force $\hat{F}$ as a function of the
dimensionless imposed approach $\hat{\delta}$ for different driving
velocities $\hat{V}$. Loading (dashed lines) and unloading (solid lines) are
performed at the same $\hat{V}$. Dotted lines connect the two branches of
the curves where the jump-in (loading phase) and jump-off (unloading phase)
instabilities occur. Results are given for a standard linear solid with $E_{0}/E_{\infty }=0.1$ and $\protect\mu _{0}=[R\Delta \protect%
\gamma ^{2}/(E_{0}^{2}\protect\varepsilon ^{3})]^{1/3}\approx 3.85$. (b) The
dimensionless contact radius $\hat{a}$ as a function of the dimensionless
force $\hat{F}$ for different driving velocities $\hat{V}$. The contact
radius is identified as the radius at which the contact pressure takes its
maximum tensile value. Loading (dashed line) and unloading (solid lines) are
performed at the same velocity. Dotted lines identify the regions where
jumping instabilities occur. Results are given for a standard linear solid with $E_{0}/E_{\infty }=0.1$ and $\protect\mu %
_{0}=[R\Delta \protect\gamma ^{2}/(E_{0}^{2}\protect\varepsilon %
^{3})]^{1/3}\approx 3.85$.}
\label{fig7}
\end{figure}

When the approximation of quasi-static loading is no longer valid and,
hence, the material is not in a completely relaxed stress state at the
beginning of unloading, the behaviour of the system significantly changes,
as shown in Fig. \textbf{\ref{fig7}}. Here, loading and unloading are performed at
the same driving velocity. Also, unloading starts right after the loading
finishes so that the substrate material does not have the time to `relax'.

With reference to Fig. \textbf{\ref{fig7}a}, the small dissipation occurring at
vanishing ($V\approx 0$) and very high ($\hat{V}>500$) velocities is
exclusively due to the adhesive hysteresis related to jumping instabilities.
The pull-off force $F_{\mathrm{PO}}$ first increases with the unloading
speed, reaches a maximum and then reduces in a similar way to the hysteresis
dissipation. This behaviour agrees with the viscoelastic modulus curves
given in Fig. \textbf{\ref{fig2}}. In fact, only in the transition region we expect
energy loss due to viscoelastic hysteresis. Therefore, at low and high
speeds, i.e., low and high frequencies of excitation, the only dissipation
is that related to the adhesion hysteresis due to the jump-on and jump-off
instabilities. In this case, we can clearly identify the two regions where
the material behaves elastically with modulus $E_{0}$ (low velocities) and $%
E_{\infty }$ (high velocities).

Figure \textbf{\ref{fig7}b}, instead, shows the contact radius, i.e., the radius
where the contact pressure takes its tensile peak, in terms of the applied
force. No difference can be identified between the loading and unloading
curves at $\hat{V}\approx 0$ and $\hat{V}>500$ as the substrate material
behaves elastically. However, at high velocities the system is much stiffer,
and the same contact area is reached at much larger forces. Moreover, in the
regions where the material is `elastic' (rubbery and glassy regions) the
contact radius significantly decreases right from the early stages of
unloading. On the contrary, at intermediate driving velocities, we observe
an initial phase where the behaviour of the contact radius is more
similar to that observed in Fig. \textbf{\ref{fig4}b}.

\begin{figure}[tbp]
\begin{center}
\includegraphics[width=12.0cm]{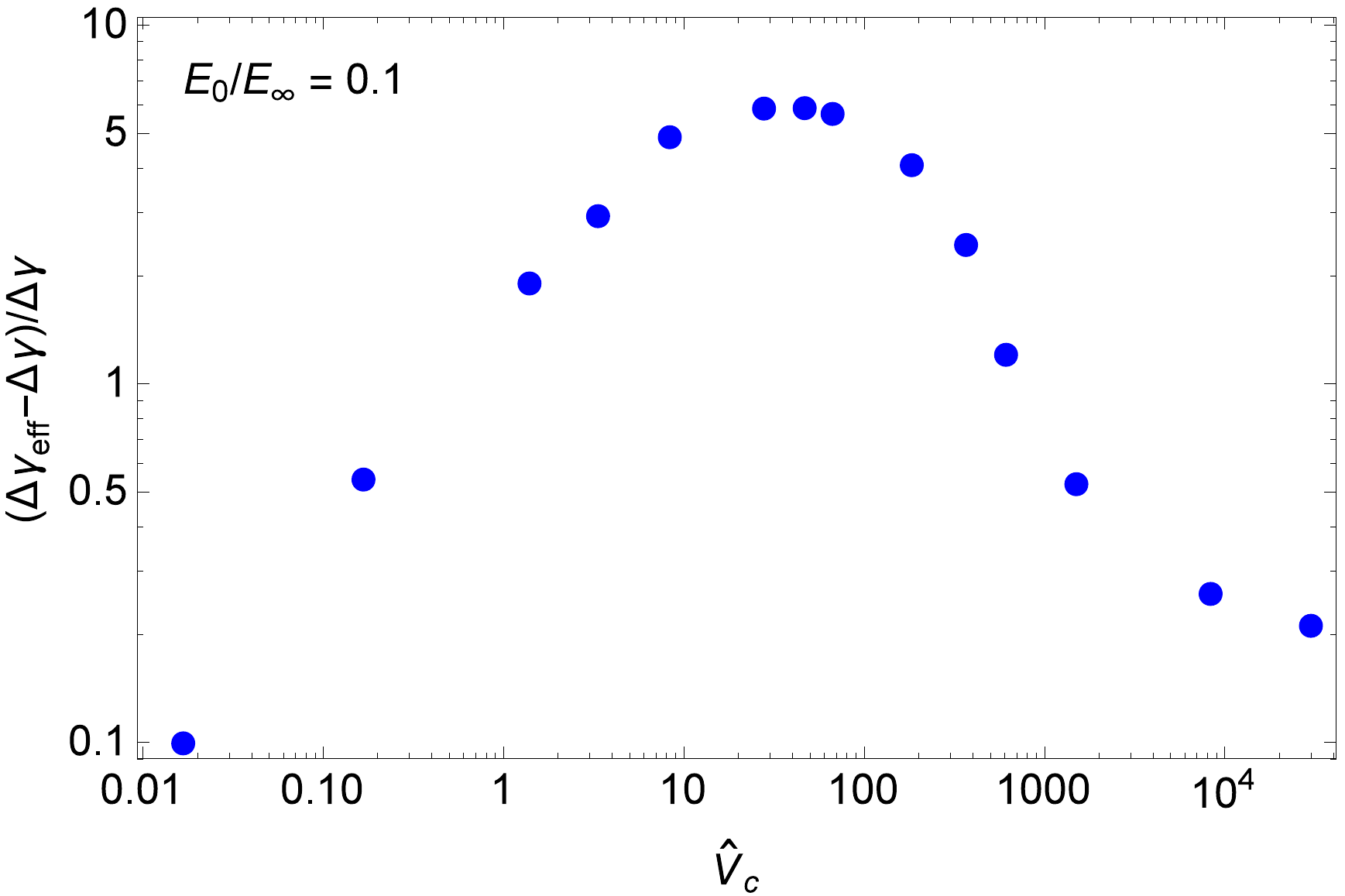}
\end{center}
\caption{The dependence of the relative increase of the effective surface
energy $\Delta \protect\gamma _{\mathrm{eff}}$ on the dimensionless
contact line velocity $\hat{V}_{\mathrm{c}}$ for a standard linear solid with $E_{0}/E_{\infty }=0.1$ and $\protect\mu %
_{0}=[R\Delta \protect\gamma ^{2}/(E_{0}^{2}\protect\varepsilon %
^{3})]^{1/3}\approx 3.85$. Results are presented in a double logarithmic
plot.}
\label{fig8}
\end{figure}

Figure \textbf{\ref{fig8}} shows the relative increase of the effective surface
energy $\Delta \gamma _{\mathrm{eff}}$ in terms of the contact line
velocity $V_{\mathrm{c}}$ as obtained from numerical simulations performed
at different driving velocities $V$. As above explained, $V_{\mathrm{c}}$ is
estimated at pull-off when we reach the maximum tensile force.\ Results are
quite different with respect to Fig. \textbf{\ref{fig4}}. Indeed, after reaching a
maximum value, $\Delta \gamma _{\mathrm{eff}}$ reduces when moving towards
high contact line velocities A bell-shape curve for the effective surface
energy has been also found in Ref. \cite{Lin2002} by numerical simulations
on two contacting viscoelastic spheres and in Ref. \cite{Gustavo} with
classical JKR-like experiments.

The velocity where $\Delta \gamma _{\mathrm{eff}}$ takes its largest value
does not correspond to the frequency at which the loss tangent is maximum.
In fact, the frequency of excitation can be roughly estimated as $\sim V/a$ and the peak of $\Delta \gamma _{\mathrm{eff}}$ is reached at about $%
V\approx 4\varepsilon /\tau $. Therefore, at pull-off being the contact
radius $\sim0.147R$, the frequency of excitation can be estimated as 
$\omega \tau \approx 0.055$, which falls in the transition region but is not
however close to the value where the loss tangent peak occurs (see Fig. \textbf{\ref{fig2}b}). A similar result is reported in Ref. \cite{francesi}, where the
fracture energy is found to reach its maximum value for low speeds "\emph{%
even if the 'liquid zone' has not emerged yet}".

Furthermore, in agreement with the considerations done in Ref. \cite%
{Israelachvili} and the calculations performed in Ref. \cite{Lin2002}, we
observe that the maximum dissipation occurs at velocity at which the time of
observation (or time scale of the process) $t_{\mathrm{o}}$ is comparable to
the characteristic time $\tau E_{\infty }/E_{0}$. Therefore, the peak of $%
\Delta \gamma _{\mathrm{eff}}$ is expected to occur close to $De\approx 1$%
, where%
\begin{equation}
De=\frac{\tau E_{\infty }/E_{0}}{t_{\mathrm{o}}}
\end{equation}
is the so-called Deborah number \cite{Reiner1964}. Indeed, we find $%
De\approx 1.35$ when the interaction time $t_{\mathrm{o}}$ is estimated as
the time between jump-on and jump-off phenomena.

\begin{figure}[tbp]
\begin{center}
\includegraphics[trim=3cm 22cm 2.2cm 0.5cm,clip=true,width=15.0cm]{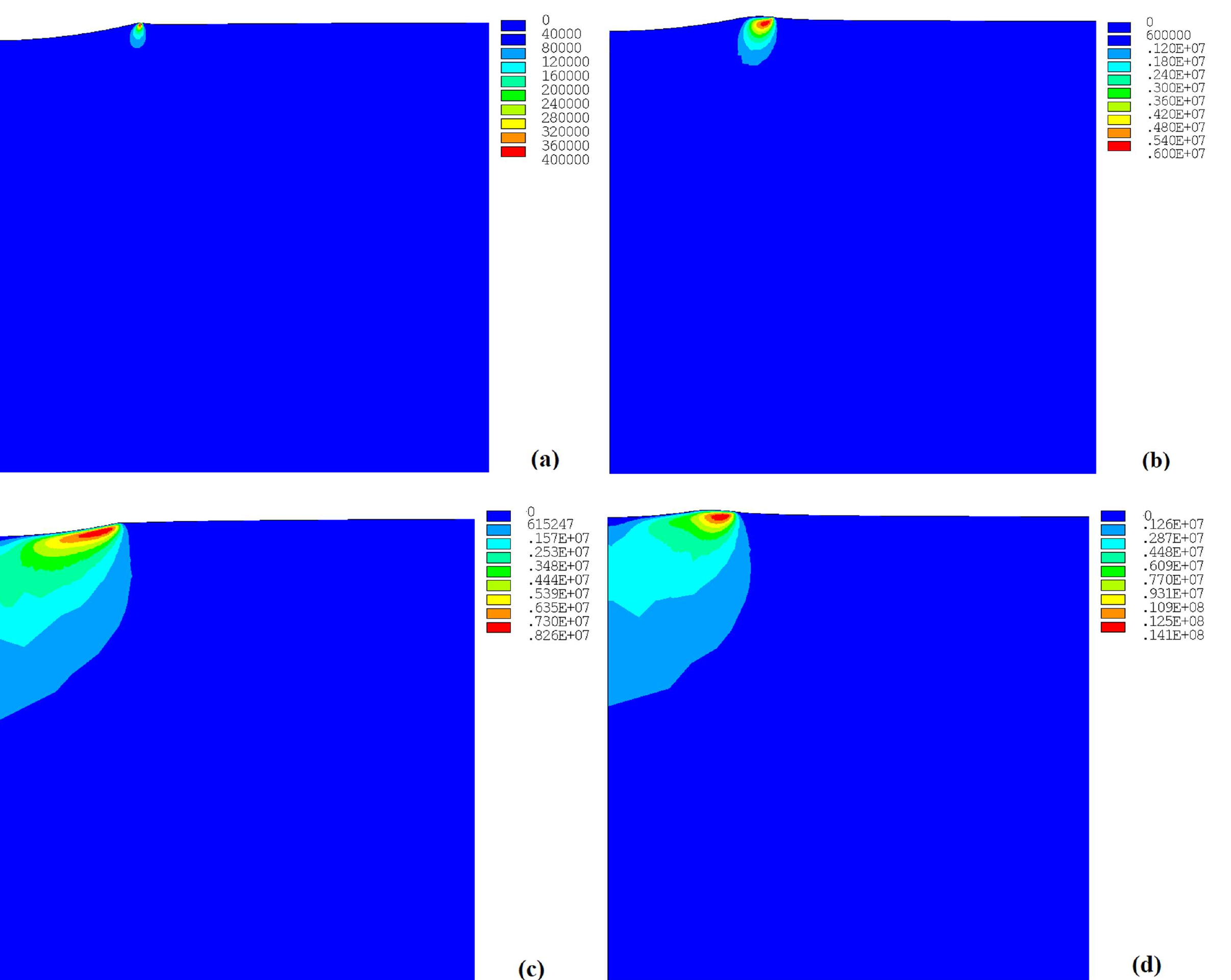}
\end{center}
\caption{The accumulated dissipated energy density at the initial phase of unloading (a-c) and at pull-off (b-d). In (a-b) loading is carried out under quasi-static conditions ($V\approx 0$) and unloading at pulling velocity $\hat{V}=1$. In (c-d) loading and unloading are performed at the same driving velocity $\hat{V}=1$. Results are obtained for a standard linear solid with $E_{0}/E_{\infty }=0.1$ and $\protect\mu %
_{0}=[R\Delta \protect\gamma ^{2}/(E_{0}^{2}\protect\varepsilon %
^{3})]^{1/3}\approx 3.85$.}
\label{fig9}
\end{figure}

The different behaviour observed in Figs. \textbf{\ref{fig4}} and \textbf{\ref{fig7}}, can
find a possible explanation in results shown in Fig. \textbf{\ref{fig9}}, where the
accumulated dissipated energy density is plotted at different times (at the
initial phase of unloading and at pull-off). Specifically, in Fig. \textbf{\ref{fig9}a-b}, loading is carried out under quasi-static conditions ($V\approx 0$) and
unloading at pulling velocity $\hat{V}=1$, while in Fig. \textbf{\ref{fig9}c-d} loading
and unloading are performed at the same $\hat{V}=1$. As expected, in the
first case, dissipation mainly occurs at the interface, near the front of
detachment. Here, therefore, the classical Gent\&Schultz assumption of
considering viscous effects localized exclusively close to the crack tip is
reasonably satisfied. In the latter case, we notice that this assumption
fails as dissipation basically occurs in the bulk.

\begin{figure}[tbp]
\begin{center}
\includegraphics[width=12.0cm]{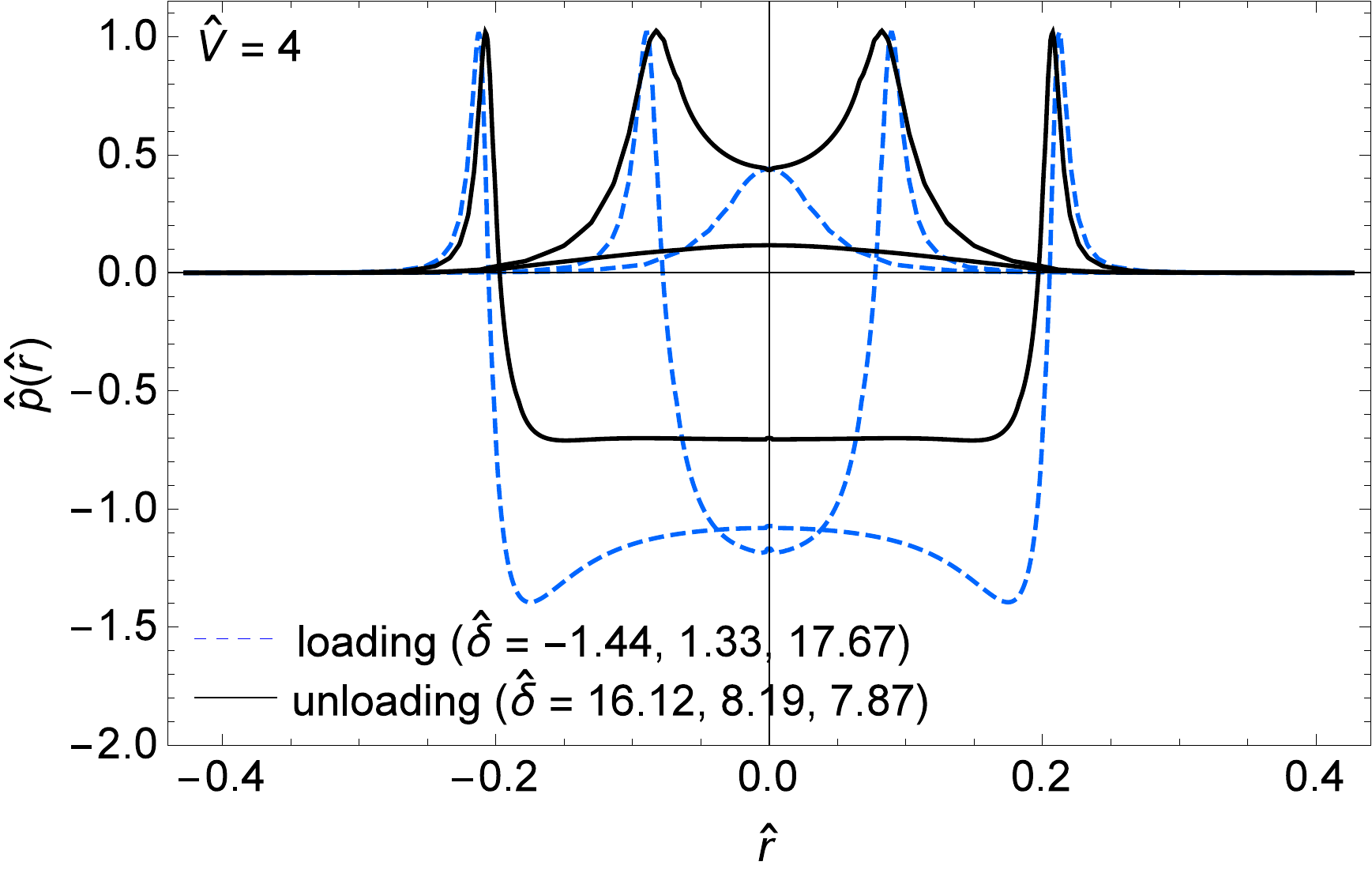}
\end{center}
\caption{The dimensionless contact pressure ($\hat{p}=p\varepsilon /\Delta 
\gamma $) distribution on the contact region for different values of the dimensionless
imposed approach $\hat{\delta}$ and $\hat{V}=4$. Dashed lines refer to the loading phase, solid lines
to the unloading one. Results are given for a standard linear viscoelastic
solid with $E_{0}/E_{\infty }=0.1$ and $\protect\mu _{0}=[R\Delta \protect%
\gamma ^{2}/(E_{0}^{2}\protect\varepsilon ^{3})]^{1/3}\approx 3.85$. }
\label{fig10}
\end{figure}

Finally, Fig. \textbf{\ref{fig10}} shows the contact pressure distribution in
different instants of the loading-unloading cycle. Again, tractions at the
edges of the contact area arise when contact is established. Moreover, we
observe that, at high penetrations, the pressure on the compressive contact
region is roughly constant.

\section{Conclusions}

In this paper, we have presented a fully deterministic numerical model for
studying the adhesive normal contact between a rigid sphere and a
viscoelastic substrate. Van der Waals adhesive interactions are described by
means of non-linear springs with a traction-gap relation according to the
Lennard-Jones law, while material viscoelasticity is modeled by a standard
linear solid with a spring in parallel with a Maxwell element constituted by
a spring in series with a dashpot.

In a first set of simulations, we have imposed quasi-static loading
conditions ($V\approx 0$) to ensure adhesive equilibrium and negligible
viscous effects. In such case, the material behaves as a soft elastic medium
(rubbery region). Unloading has been instead performed for different driving
velocities. We found that, as unloading starts from a completely relaxed
state of the material, a progressive enhancement of the effective surface
energy with the contact line velocity is found up to an asymptotic value.
This is reached for high contact line velocities that correspond to
excitation frequencies falling in the glassy region of the material, where
it behaves as a stiff elastic medium. Numerical data of $\Delta \gamma _{%
\mathrm{eff}}(V_{\mathrm{c}})$ are found in good agreement with the
theoretical predictions of two crack's propagation theories, namely the
cohesive-zone model proposed by Greenwood \cite{Greenwood2004} on the basis
of Schapery's findings, and the energetic approach developed by Persson \&
Brenner \cite{PB2005}.

In a second set of simulations, we have performed loading and unloading at
the same driving velocity. No dwell time has been waited before unloading.
In such case, the trend of the effective surface energy with the contact
line velocity is described by a bell-shaped function. The maximum of the
bell is found for exciting frequency lower than the frequency that
maximize the tangent loss of the viscoelastic modulus in agreement to
previous studies. Moreover, the peak of dissipation is found close to $De=1$%
. We have also discussed how the Gent\&Schultz assumption works in this case
showing that viscous dissipation is not always localized along the contact
perimeter but it may occur in the bulk material. In such case, adhesion models based on linear elastic fracture mechanics are expected to be less accurate.

\acknowledgements{L.A. and G.V. acknowledge support from the Italian Ministry of Education, University and Research (MIUR) under the program "Departments of Excellence" (L.232/2016). Moreover, this work was supported by the project "FASTire (Foam Airless Spoked Tire): Smart~Airless Tyres for Extremely-Low Rolling Resistance and Superior Passengers Comfort" funded by the Italian MIUR Progetti di Ricerca di Rilevante Interesse Nazionale (PRIN) call 2017---grant n. 2017948FEN.}

\end{document}